\documentclass[a4paper,11pt]{article}
\usepackage{pos}
\usepackage{cancel}
\usepackage[normalem]{ulem}
\usepackage{float}
\usepackage{autobreak}
\usepackage{graphicx}
\usepackage{appendix}
\usepackage{amsfonts}
\usepackage{comment}
\usepackage{multirow}
\usepackage{color,hyperref}
\usepackage[T1]{fontenc}
\usepackage{tocloft}
\usepackage[normalem]{ulem}
\usepackage{subcaption}
\usepackage{array}
\usepackage{enumitem}
\usepackage{extarrows}
\usepackage{mathtools}
\usepackage{slashed}

\newcommand{\nn}{\nonumber\\}

\def\D#1{{{\cal D}_{#1}}}

\newcommand{\df}{{\rm d}}
\def\Dm1{{{\delta(1-z)}}}

\def\g0#1DY{{g_{0#1}^{DY}}}

\def\LogmW1{{{\ln (1-\omega)}}}

\newcommand{\overbar}[1]{\,\overline{\!{#1}}}
\newcommand{\Nbar}{\overbar{N}}
\newcommand{\gbar}{\overbar{g}}
\newcommand{\as}{a_s}
\newcommand{\muf}{\mu_F}
\newcommand{\mur}{\mu_R}

\newcommand{\zb}{{\bar{z}}}

\newcommand{\NLL}{\overline{\text{NLL}}}

\newcommand{\eq}[1]{Eq.\ (\ref{#1})}
\newcommand{\fig}[1]{Fig.\ \ref{#1}}

\newcommand{\sect}[1]{Section\ \ref{#1}}

\title{Threshold resummation for gluon fusion $ZH$ production at the LHC}
\author[a,b]{Goutam Das}
\author[c]{Chinmoy Dey}
\author[d]{M. C. Kumar}
\author*[e]{Kajal Samanta}

\affiliation[a]{Institut f{\"u}r Theoretische Teilchenphysik und Kosmologie, RWTH Aachen University, \\D-52056 Aachen, Germany}
\affiliation[b]{Department of Physics, Indian Institute of Technology Kanpur, Kanpur-208016, India}

\affiliation[c]{Theoretical Physics Division, Physical Research Laboratory,\\ Navrangpura, Ahmedabad 380009, India}

\affiliation[d]{Department of Physics, Indian Institute of Technology Guwahati,\\ Guwahati-781039, Assam, India}

\affiliation[e]{ Institute for Particle Physics Phenomenology, Durham University,\\ Durham DH1 3LE, United Kingdom}

\emailAdd{gkdgoutam@iitk.ac.in}
\emailAdd{chinmoy@prl.res.in}
\emailAdd{mckumar@iitg.ac.in}
\emailAdd{kajal.samanta@durham.ac.uk}

\abstract{We present precise results for the $ZH$ production cross-section and invariant mass distribution at the LHC, taking into account the effects of the leading and sub-leading soft gluons. We improve both quark-initiated and gluon-initiated subprocesses through threshold resummation within the QCD framework.}

\FullConference{17th International Symposium on Radiative Corrections: Applications of Quantum Field Theory to Phenomenology (RADCOR2025)\\
5-10 October 2025\\
Puri, India\\}

\begin{document}
{\flushright{TTK-26-02, IPPP/26/01}}
\maketitle

\section{Introduction}\label{sec:introduction}
The discovery of the Higgs boson at the Large Hadron Collider 
(LHC) is a significant milestone in particle physics. Understanding the Higgs is very important for both the Standard Model (SM) and beyond-the-Standard-Model (BSM) physics. Many BSM scenarios predict weak couplings to the Higgs boson at collider energies in the TeV range. These effects can be seen as deviation from the SM prediction. Furthermore, it is also important to constrain the sign of the top-quark Yukawa coupling and examine its $CP$ structure \cite{Englert:2013vua, Hespel:2015zea, Goncalves:2015mfa}.  Keeping in mind the importance 
of this process, the ATLAS and CMS experiments are actively conducting precise measurements for this 
process \cite{CMS:2024fkb, ATLAS:2024yzu}. Precise theoretical predictions are thus essential to complement the improved experimental 
results for this channel.

At the LHC, $ZH$ is primarily produced through the  Drell-Yan (DY) type subprocess 
$(q\bar{q} \rightarrow Z^* \rightarrow ZH)$ at the leading order (LO) in the strong coupling ($\alpha_s$) expansion. The 
higher-order quantum chromodynamics (QCD) corrections for this channel thus closely follow the DY process and have been 
computed up to next-to-next-to-leading order (N2LO) 
\cite{Campbell:2016jau,Ferrera:2017zex,Harlander:2018yio} 
and recently to next-to-next-to-next-to-leading order (N3LO) \cite{Baglio:2022wzu} accuracy. Compared to the DY process, the $ZH$ production process receives additional QCD contributions starting from N2LO. One such contribution arises from Feynman diagrams in which the Higgs boson is radiated off a massive top-quark loop in the quark–antiquark annihilation subprocess. The impact of this contribution has been estimated in \cite{Brein:2011vx} to be below $3\%$. Another class of subprocesses that first appears at N2LO, i.e. at ${\cal O}(\alpha_s^2)$, is the gluon fusion channel for 
$ZH$ production. Although the gluon fusion subprocess is suppressed by two powers of the strong coupling constant compared to the quark–antiquark annihilation channel, this suppression is largely compensated by the sizable gluon luminosity at the LHC. Consequently, from N2LO onwards, the gluon fusion contribution becomes phenomenologically significant. Several efforts have been made to further improve the accuracy of the gluon-initiated subprocess by calculating its NLO ($\mathcal{O}(\alpha_s^3)$) contribution. Recent development of such fixed-order computation can be found in \cite{ Bellafronte:2022jmo, Degrassi:2022mro, Wang:2021rxu, Chen:2020gae, Chen:2022rua}. 

The fixed-order results for this subprocess still suffer from large threshold logarithms arising from soft gluons emission. These large logarithms need to be resummed to have reliable predictions. The size of the NLO corrections indicates that this subprocess will receive significant contributions from the threshold logarithms similar to the Higgs case. Resummation of these large soft-virtual (SV)  logarithms is well established in the literature \cite{Sterman:1986aj, Catani:1990rp, Ravindran:2005vv} 
%
%\cor{ADDED: \cite{Kidonakis:1997gm,Kidonakis:2003tx,Kidonakis:2005kz} }
%
and they have been applied to many colorless processes, leading to improved predictions for inclusive cross sections and invariant mass distributions. Recently, efforts were made to incorporate also the next-to-soft (NSV) threshold effects
\cite{vanBeekveld:2019cks,AH:2020iki}. The main focus of this paper is to improve the gluon fusion $ZH$ process by incorporating soft and next-to-soft gluon resummation 
for both the total cross section and the invariant mass distribution of the $ZH$ pair.

The article is organized as follows: In \sect{sec:theory}, we introduce the key theoretical formulas for performing SV and NSV resummation up to the necessary order. In \sect{sec:numerics}, we provide a phenomenological 
study for the gluon fusion subprocess, combining it with the DY-type contributions to present complete results for $pp$ 
collisions at $\mathcal{O}(\alpha_s^3)$ accuracy.  Finally, we conclude in \sect{sec:conclusion}.

\section{Theoretical framework}\label{sec:theory}
In QCD factorization, the hadronic cross section for $ZH$ production in $pp$ collisions can be written as 
\begin{align}\label{eq:had-xsect}
Q^2 \frac{\df \sigma}{\df Q^2}  
=  
\sum_{a,b}\int_0^1 \df x_1\int_0^1 \df x_2 \,\,f_{a}(x_1,\mu_F^2)\,
f_{b}(x_2,\mu_F^2) 
\int_0^1 \df z~ \delta \left(\tau-zx_1 x_2\right)
Q^2 \frac{\df \widehat\sigma_{ab}(z,\muf^2)}{\df Q^2} \, ,
\end{align}
where $f_{a,b}$ are the parton distribution functions (PDFs) for the parton $a,b$ in the incoming hadrons and $\widehat\sigma_{ab}$ is the partonic coefficient function.  
The factorization scale is $\muf$, $Q$ is the invariant mass of the final state, $\tau$ and $z$ are the hadronic and partonic threshold variables, respectively, and are defined as $\tau=Q^2/S$, $ z= Q^2/\widehat{s}$,  where $S$ and $\widehat{s}$ are the hadronic and partonic center-of-mass energy. The partonic coefficient funtion $\widehat\sigma_{ab}$ can be decomposed into two parts; singular parts and regular parts. The singular part contains all the singular contributions with respect to the partonic threshold variable $z$ and the regular part is finite. The decomposition can be written as,
\begin{align}\label{eq:PARTONIC-DECOMPOSE}
Q^2 \frac{\df \widehat\sigma_{ab}(z)}{\df Q^2}
& =
\widehat{\sigma}^{(0)}_{ab}(Q^2) \Big( 
\delta_{a {b}}\Delta_{ab}^{\rm SV}\left(z\right) 
+ \Delta_{ab}^{\rm REG}\left(z\right)
\Big) \,.
\end{align}
where all the scale dependencies are implicit. The $\delta_{a {b}}$ shows that the singular part gets contribution only through diagonal channel. 
In general, the regular and singular part can be written as a series expansion of the strong coupling constant as,
\begin{align}\label{eq:SVREG-Expansion}
\Delta^{\rm SV}_{ab}(z) 
&= \sum_{n=0}^{\infty}\as^n(\mur^2) ~\delta_{ab}
\left( 
\Delta^{(n)}_{\delta} ~\delta(\zb) + 
\sum_{k=0}^{2n-1}  \Delta^{(n)}_{{\cal D}_k} ~{\cal D}_k(\zb) 
\right)\,, ~ 
\text{with } ab \in \{gg, q\bar{q} \} \,,
\\
\Delta^{\rm REG}_{ab}(z) 
&= 
\Delta^{\rm NSV}_{ab}(z) 
+
\sum_{n=0}^{\infty}\as^n(\mur^2)
%\left( 
\Delta_{ab}^{(n)}(z)
= \sum_{n=0}^{\infty}\as^n(\mur^2)
\left( 
\sum_{k=0}^{2n-1} \Delta_{ab,{\ln}_k}^{(n)} ~\ln^k(\zb) 
+
\Delta_{ab}^{(n)}(z)
\right)\,,
\nonumber
\end{align}
where $\zb = (1-z)$, $\delta(\zb)$ is the Dirac delta distribution and ${\cal D}_k(\zb) \equiv \left[\ln^k(\zb)/\zb\right]_+$ are 
the plus distributions.
The $\Delta_{ab,{\ln}_k}^{(n)} $ term are the NSV coefficients which get contributions from both diagonal as well as off-diagonal channels.
The SV part of the partonic coefficient has a universal structure which gets contributions from the underlying hard form factor, mass factorization kernels
%\cite{Moch:2004pa,Vogt:2004mw} 
and soft radiations.
%\cite{Ravindran:2005vv,
%Ravindran:2006cg,
%Sudakov:1954sw,
%Mueller:1979ih,
%Collins:1980ih,
%Sen:1981sd}. 
The plus distribution function in singular part of partonic cross section shows that in the threshold limit $z \to 1$, it can give a large contribution. To have a reliable theoretical prediction, one needs to resum these lagre logarithms to all orders.
Threshold resummation is generally performed in the Mellin $N$-space,
where plus distributions become simple logarithms in the Mellin variable ($N$). The threshold limit $z \to 1$
translates into the $N \to \infty$ limit in the Mellin space. The formalism for the SV  threshold resummation is known for quite long time in the literature.
Recently a formalism has been proposed 
\cite{AH:2020iki, vanBeekveld:2021hhv} to also resum the NSV logarithms 
arising out of the diagonal channel. 
The formalism is built upon a factorisation framework that includes a process-dependent hard function, mass factorisation kernels, and a soft function, as detailed in \cite{AH:2020iki}. The structure can be explicitly verified at NLO, the higher-order coefficients are determined using the known fixed-order results for the Higgs \cite{Mistlberger:2018etf} and DY \cite{Duhr:2020seh} production processes.

The Born coefficient in Eq.~\ref{eq:had-xsect} in the exact theory contains Feynman diagrams involving triangle and box diagrams 
with heavy top-quark mass dependence. It can be expressed in terms of helicity amplitudes for triangle and box diagrams as,
\begin{align}\label{eq:LO-EXACT}
\widehat{\sigma}^{(0)}_{gg}(Q^2) =&
\left( \frac{ \sqrt{\pi} \as(\mur^2) \alpha}{16 s_w^2} \right)^2 \frac{1}{8\widehat{s}^2}
\int \df \widehat{t} \sum_{\lambda_g,\lambda^{'}_g,\lambda_Z} \left|
{\mathcal M}_{\lambda_g\lambda^{'}_g\lambda_Z}^\triangle
+{\mathcal M}_{\lambda_g\lambda^{'}_g\lambda_Z}^\Box\right|^2 \,,
\end{align}
where $\widehat{s}$ and $\widehat{t}$ are the partonic Mandelstam variables.
The expression for these helicity amplitudes can be found in \cite{Kniehl:2011aa}. 
In case of EFT, the Born coefficient function can be found in \cite{Altenkamp:2012sx}.

The partonic SV and NSV coefficients in the Mellin space can be organized as follows,
\begin{align}\label{eq:resum-partonic}
	\frac{1}{\widehat{\sigma}^{(0)}_{ab}(Q^2)}Q^2\frac{\df \widehat{\sigma}_{N,ab}^{\overline{\rm N{\it n}LL}}}{\df Q^2}
	= \int_0^1 \df z ~ z^{N-1} \left(  \Delta^{\rm SV}_{ab}(z) + \Delta^{\rm NSV}_{ab}(z)  \right)
	\equiv g_{0}(Q^2) \exp \left( G^{\rm SV}_N + G^{\rm NSV}_N \right)\,. 
\end{align}
The factor $g_{0}$ is independent of the Mellin variable and contains process-dependent information. This factor can be written as a series expansion of the strong coupling constant as, 
\begin{align}
	g_0(Q^2) = 1+\sum_{n=1}^{\infty} \as^n(\mur^2) ~g_{_{0n}}(Q^2) \,.  
\end{align}
The leading threshold-enhanced large logarithms and the next-to-soft logarithms in Mellin space are resummed through 
the exponents $G^{\rm SV}_{N}$ and $G^{\rm NSV}_{N}$ respectively.
The resummed accuracy is determined through the 
successive terms from the exponent $G_N$, which takes the form,
\begin{align}\label{eq:GN}
G^{\rm SV}_N &= 
	\ln (\Nbar) ~g_1(\omega)
        + \sum_{n=1}^{\infty} \as^{n-1}(\mur^2)~ g_{n+1}(\omega) \,,
 \nn
G^{\rm NSV}_N &= 
        \frac{1}{N}
        \sum_{n=0}^{\infty} \as^{n}(\mur^2)~ \left( \gbar_{n+1}(\omega) + \sum_{k=0}^{n} h_{nk}(\omega) \ln^k \Nbar \right)\,,       
\end{align}
where $\Nbar = N \exp (\gamma_E)$ with $\gamma_E$ being the Euler–Mascheroni constant and $\omega = 2 \beta_0 \as(\mur^2) \ln \Nbar$.
The first term in the expansion of $G^{\rm SV}_N$ corresponds to the leading logarithmic (LL) accuracy, whereas the inclusion of successive terms defines higher accuracy.
These coefficients ($g_n$) are universal and only depend on the partonic flavors (quark or gluon). To obtain $\overline{\rm LL}$ accuracy in NSV resummation, one has to consider NSV 
exponents $\overline{g}_1$ and $h_{00}$ in addition to the LL terms. 
Similarly, for higher accuracies, one has to consider the next terms in the expansion of $G_N$. All the ingredients for NLL($\NLL$) can be found in \cite{Das:2025wbj}.

The resummed result in \eq{eq:resum-partonic} must finally be matched with the available fixed-order results to avoid double-counting of SV(NSV) logarithms.
The matching with the fixed order is usually performed using the \textit{minimal prescription} \cite{Catani:1996yz} and the matched results can be written as,
\begin{align}\label{eq:MATCHING}
	Q^2\frac{\df \sigma^{\rm N{\it n}LO+\overline{N{\it n}LL}}_{ab}}{\df Q^2}
	=&
	Q^2 \frac{\df {\sigma}^{\rm N{\it n}LO}_{ab} }{\df Q^2}
	+
	%\sum_{a,b \in \{q,\bar{q}\}}
        \sum_{ab \in \{gg, q\bar{q}\} }
        \widehat{\sigma}^{(0)}_{ab}(Q^2)
	\int_{c-i\infty}^{c+i\infty}
	\frac{\df N}{2\pi i}
	\tau^{-N}
	f_{a,N}(\muf)
	f_{b,N}(\muf)
	\nn
	&\times 
	\Bigg( 
		Q^2\frac{\df \widehat{\sigma}_{N,ab}^{\overline{\rm N{\it n}LL}}}{\df Q^2} 
		- 	
		Q^2\frac{\df \widehat{\sigma}_{N,ab}^{\overline{\rm N{\it n}LL}}}{\df Q^2} \Bigg|_{\rm tr}
	\Bigg) \,.   
\end{align}
The SV resummation matching procedure is very similar to NSV matching.
The $f_{a,N}$ are the Mellin-space 
PDFs similar to the partonic coefficient in 
\eq{eq:resum-partonic} and can be approximated 
by directly using the $z$-space PDF following 
\cite{Catani:2003zt}.
In the next section, we study the impact of SV and NSV resummation on the gluon subprocess at the LHC.

\section{Numerical results}\label{sec:numerics}
\begin{figure}
\centerline{
\includegraphics[width=7.5cm, height=7.5cm]{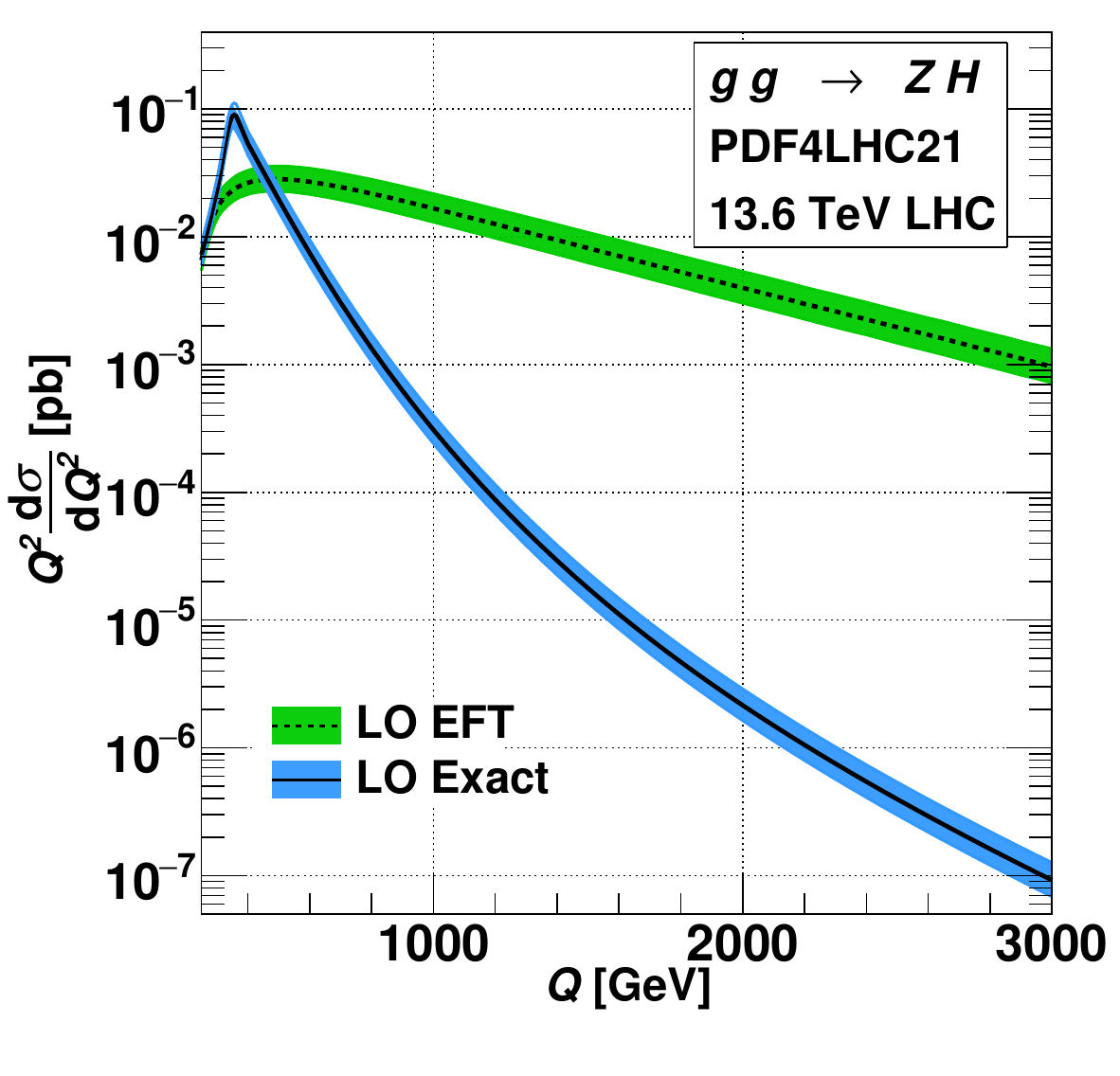}
}
\vspace{-2mm}
\caption{\small{The comparison of invariant mass distribution of $ZH$ in the gluon fusion channel at LO for EFT and Exact theories 
	at $13.6$ TeV LHC. The bands correspond to the theoretical uncertainty using seven-point scale variation. }}
\label{fig:fo_kfac_eft_gg}
\end{figure}

In this section, we present numerical results for $ZH$-associated production at the LHC. Our default choice of center-of-mass energy is $13.6$ TeV. We use 
{\tt PDF4LHC21\_40} \cite{PDF4LHCWorkingGroup:2022cjn} parton distribution functions (PDFs) throughout, as provided by 
{\tt LHAPDF}. In all these cases, the central set is the standard choice. The strong coupling is 
provided through the {\tt LHAPDF} routine. The fine structure constant is taken as $\alpha \simeq 1/127.93$. The masses 
of the weak gauge bosons are set to be $M_Z = 91.1880$ GeV and $M_W = 80.3692$ GeV with 
the corresponding total decay widths of the $Z$ boson, $\Gamma_Z = 2.4955$ GeV. The Weinberg angle is then given by 
$\text{sin}^2\theta_\text{w} = (1 - m_W^2/m_Z^2)$.
This corresponds to the weak coupling $G_F \simeq 1.2043993808\times 10^{-5} \text{ GeV}^{-2}$.
The mass of the Higgs boson is set to $M_H = 125.2$ GeV. 
For the quark masses, we use the top-quark pole mass $m_t = 172.57$ GeV and  pole mass of the bottom quark $m_b = 4.78$ GeV. The unphysical scales ($\mur$ and $\muf$) are set to the invariant mass ($Q$) of the $ZH$ pair. The scale uncertainties 
are estimated by using conventional seven-point scale variation.
To estimate the impact of the higher-order corrections from FO and resummation, we define the following ratios of the cross sections:
\begin{align}
	{ K}_{nm}
	= 
	\frac{\sigma^{\text{N}{\it n}\text{LO}}}{\sigma_c^{\text{N}{\it m}\text{LO}}} 
	,
	{R}_{nm} 
	= 
	\frac{\sigma^{\text{N}{\it n}\text{LO} + \text{N}{\it n}\text{LL}}}{\sigma_c^{\text{N}{\it m}\text{LO}}} \text{ and }
	{ \overbar{R}}_{nm} 
	= 
	\frac{\sigma^{\text{N}{\it n}\text{LO} + \overline{\text{N}{\it n}\text{LL}}}}{\sigma_c^{\text{N}{\it m}\text{LO}}} \,.
	\label{eq:ratio}
\end{align}
%\subsection{Invariant mass distribution}
First, we study the invariant mass distribution for $ZH$ in the effective and full theories. For the production cross section, these two theories differ by around $30\%$. However, for the invariant mass distribution, the difference is prominent as shown in \fig{fig:fo_kfac_eft_gg}. Below the top threshold, the two theories are comparable, which is not true anymore near the top threshold and beyond. Because higher-order QCD computations with full top-mass dependence are very complicated, we use Born-improved NLO for our fixed-order results. In this born-improved NLO, we compute the NLO k-factor in the effective theory and then multiply it by the exact top mass-dependent LO order results to have NLO results. As the top mass has less impact on the higher-order K-factor, with this approach, we can correctly produce the shape of the invariant mass distribution.

\begin{figure}[ht!]
\centerline{
\includegraphics[width=7.5cm, height=7.5cm]{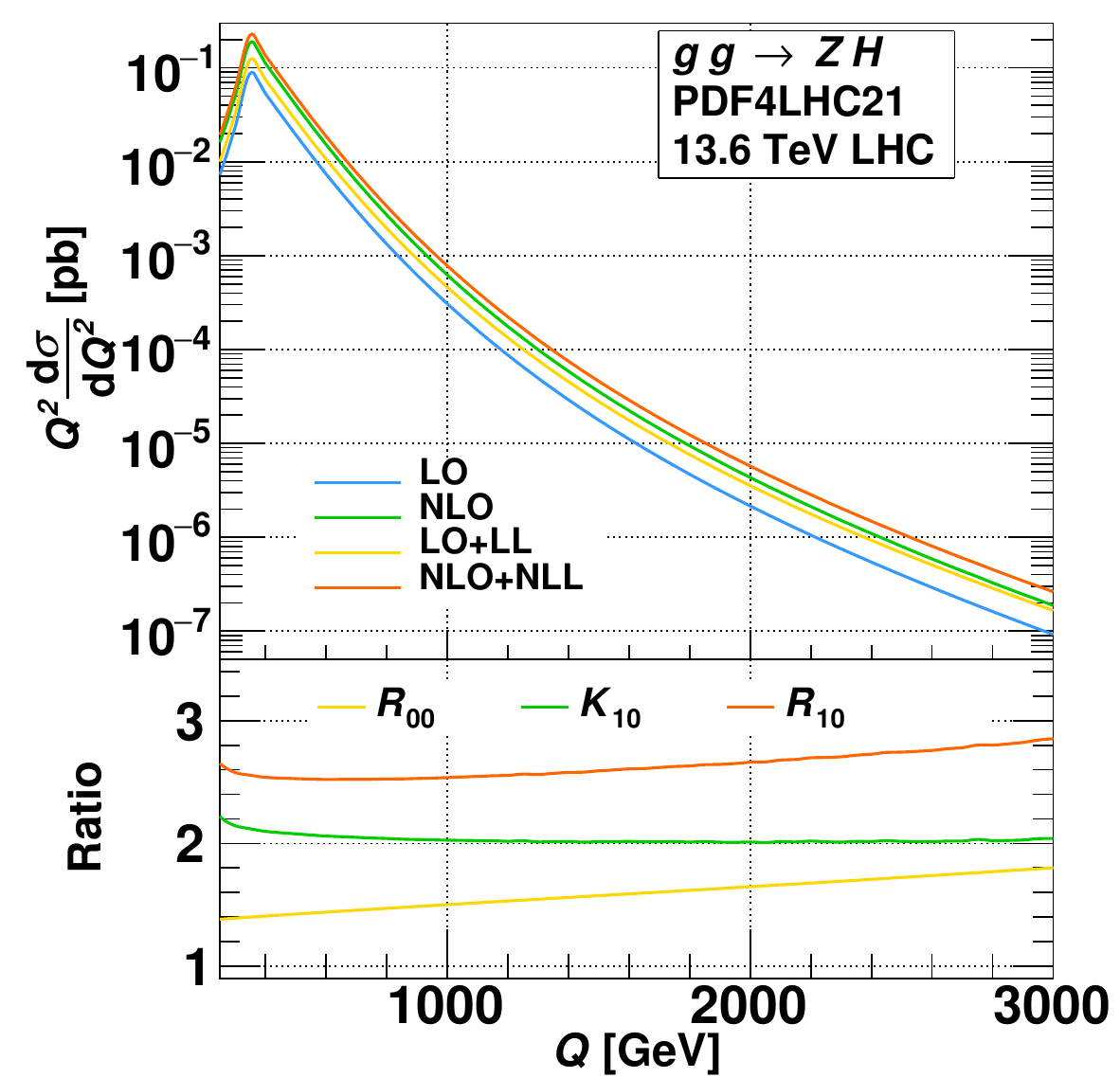}
\includegraphics[width=7.5cm, height=7.5cm]{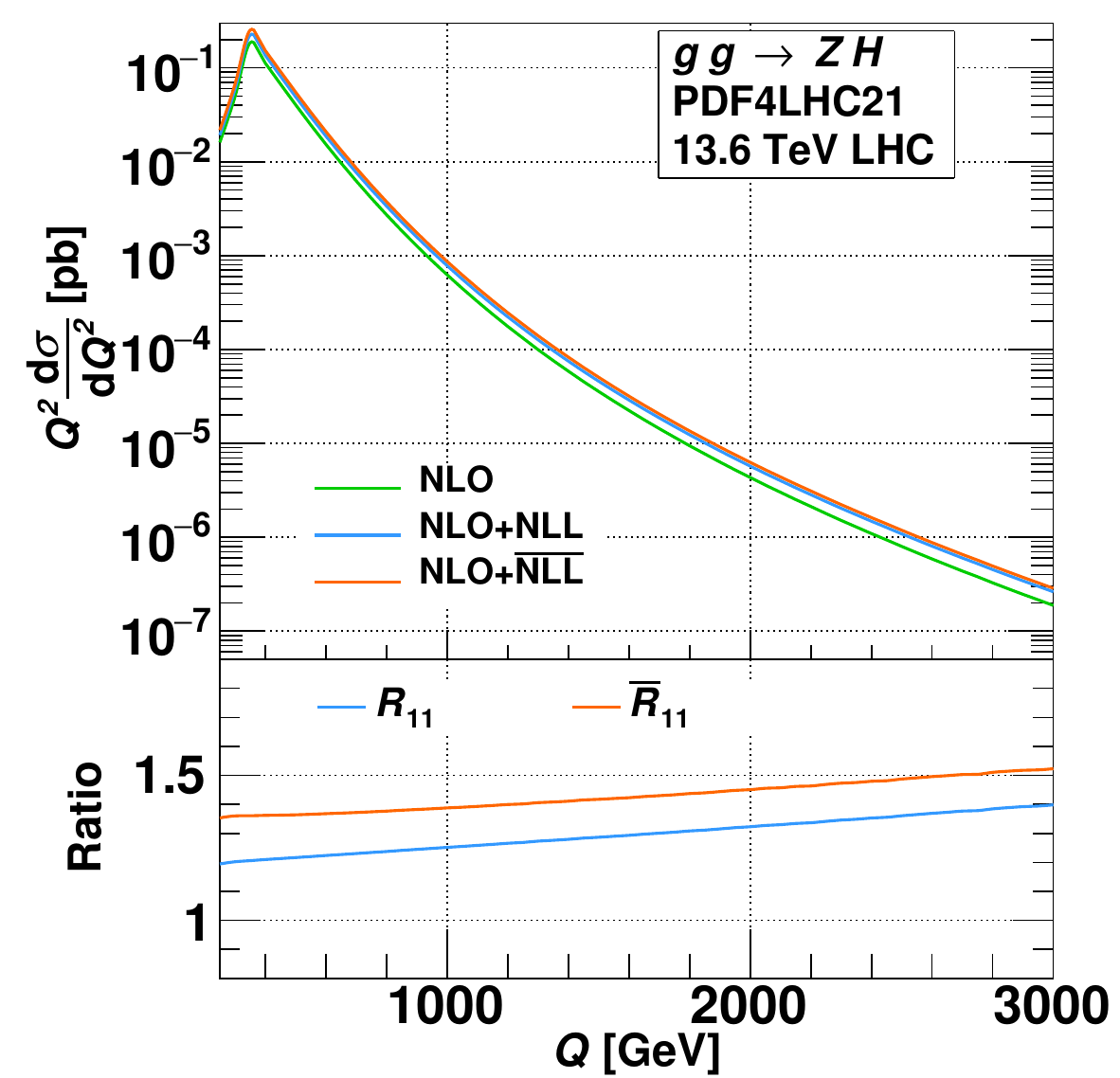}
}
\vspace{-2mm}
\caption{\small{Comparison between the Born-improved fixed order and resummed results. All the ratios as 
	defined in \eq{eq:ratio} to quantify the enhancement. The left panel is for fixed order and SV resummation and the right panel is for fixed order, SV resummation and NSV resummation.}}
\label{fig:SV_NSV_gg}
\end{figure}

In the left panel of \fig{fig:SV_NSV_gg}, we compare the SV resummed results with those of the fixed-order up to NLO(NLL) 
level, and we observe the expected behavior of better perturbative convergence for the SV resummed series. The LO+LL correction provides an additional contribution 
of $80\%$ of the LO, particularly in the high invariant mass region ($Q=3000$ GeV). This behaviour indicates that the soft-gluon effects are  dominant for this process in the high invariant mass region. The NLO+NLL correction also shows a significant enhancement 
(through the $R_{10}$ factor), reaching around $2.8$ times the LO, particularly in the high invariant mass region. In the right panel of \fig{fig:SV_NSV_gg}, we compare the NLO(NLL) results 
with $\NLL$ results. Through the ratios $R_{11}$ and $\overbar{R}_{11}$ in the bottom panel, we observe 
that NLO+NLL corrections account for about $20\%$  of NLO in the low invariant mass region and rise to approximately 
$40\%$ in the high invariant mass region. The NSV resummation at NLO+$\NLL$, on the other hand, contributes an additional $12\% - 15\%$ correction over NLO+NLL across most of the invariant mass region.

\begin{figure}[ht!]
\centerline{
\includegraphics[width=7.5cm, height=7.5cm]{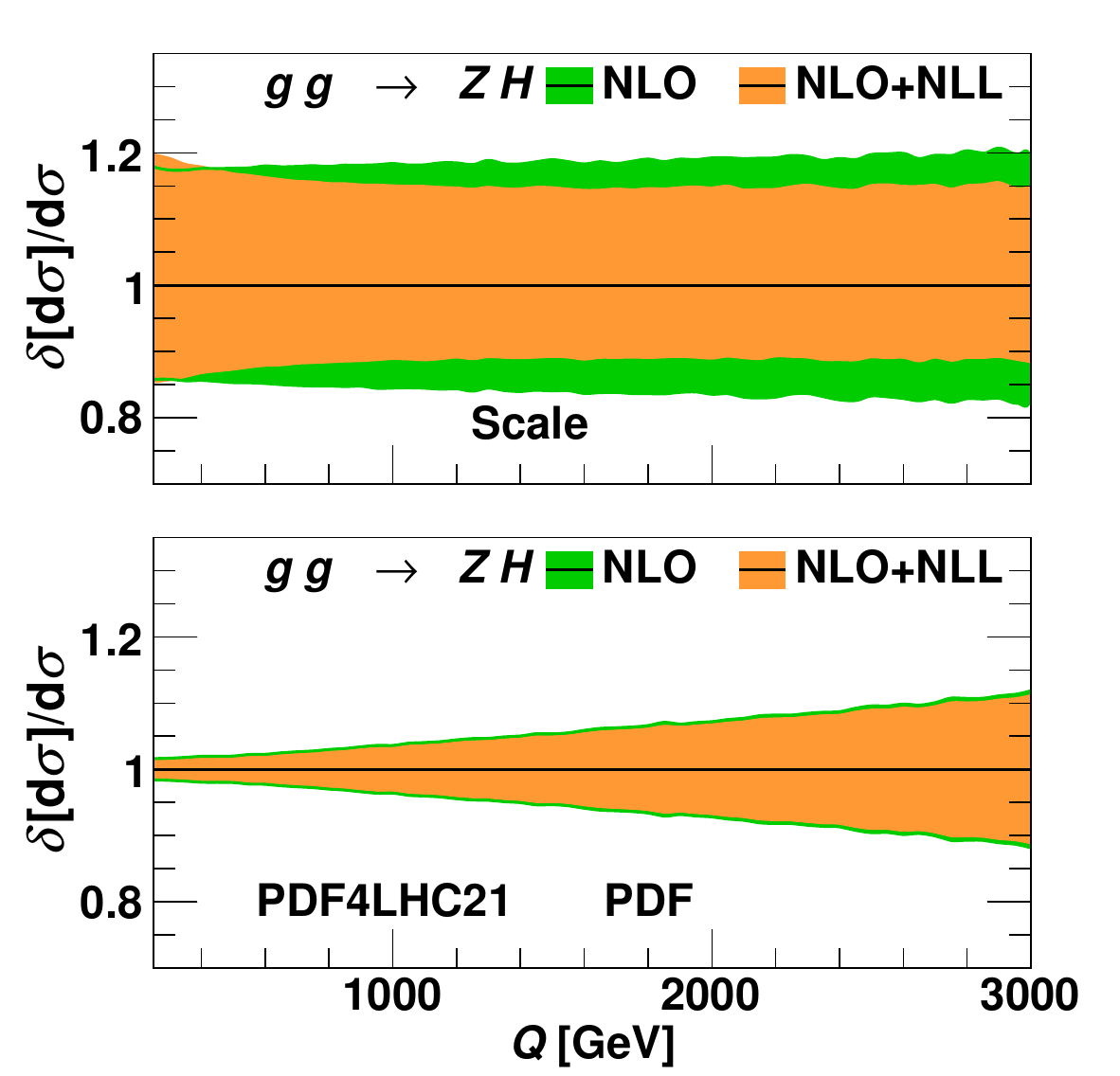}
\includegraphics[width=7.5cm, height=7.5cm]{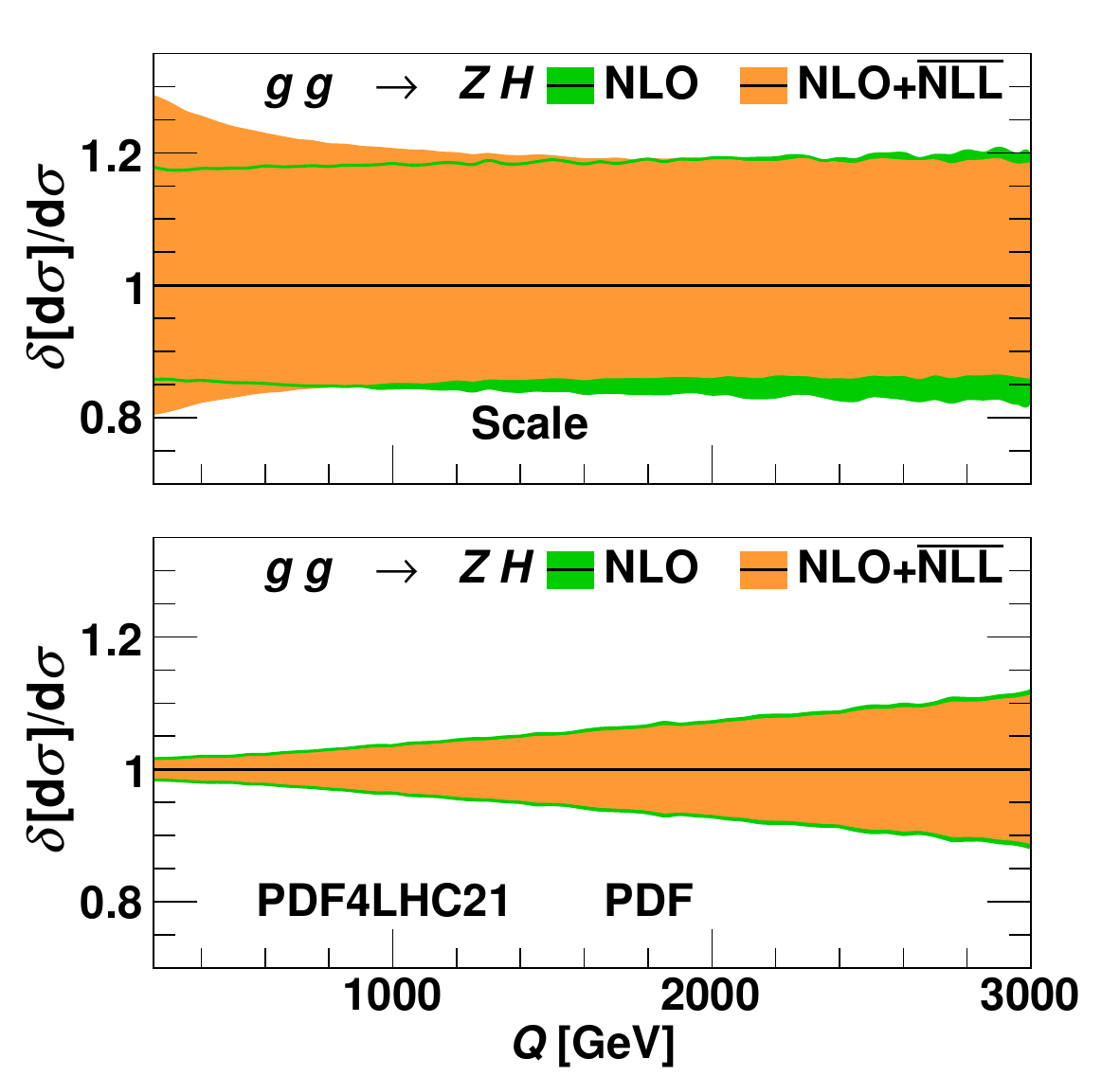}
}
\vspace{-2mm}
\caption{\small{Comparison of 7-point scale (upper panel) and the PDF uncertainty (lower panel) of gluon fusion 
	$ZH$ production for Born-improved SV (left) and NSV resummation (right) against their corresponding 
	fixed order results at $13.6$ TeV LHC.}}
\label{fig:scale_gg}
\end{figure}

We have studied the different sources of uncertainty in our prediction and presented in \fig{fig:scale_gg}. 
We observe that the scale uncertainties at NLO are around $20\%$, which gets reduced to around $15\%$ in NLO+NLL distribution.
On the other hand, NSV resummation at NLO+$\NLL$ does not show similar scale behavior. In the high invariant mass region, it marginally improves the scale uncertainty 
over the NLO results. In the bottom panels of \fig{fig:scale_gg}, we compare the PDF uncertainties for both SV and NSV resummed cases against the fixed order. The SV and NSV resummed results show marginal improvements in PDF uncertainties by around $0.8\%$ over the NLO.

\begin{figure}[ht!]
\centerline{
\includegraphics[width=7.5cm, height=7.5cm]{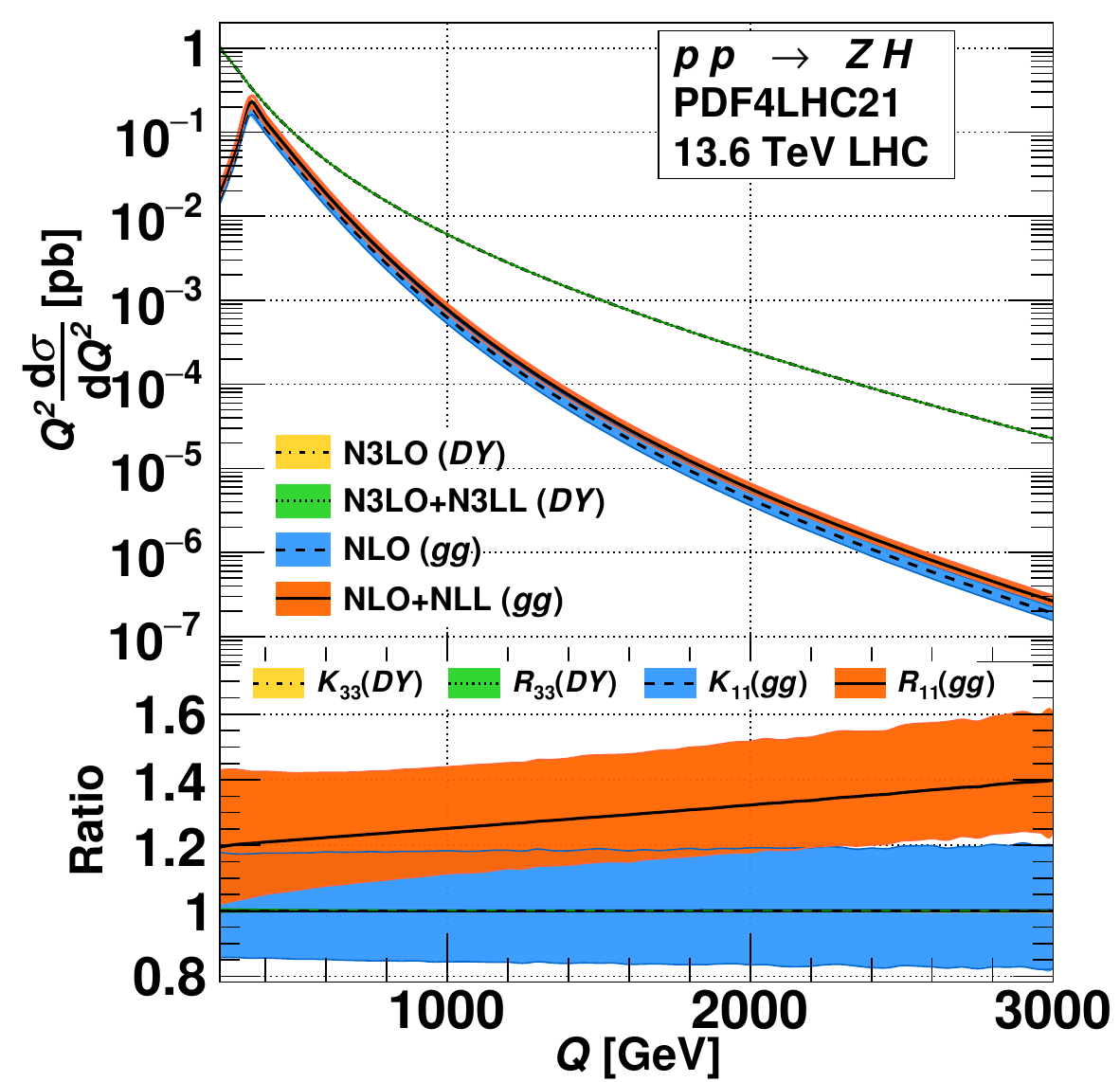}
\includegraphics[width=7.5cm, height=7.5cm]{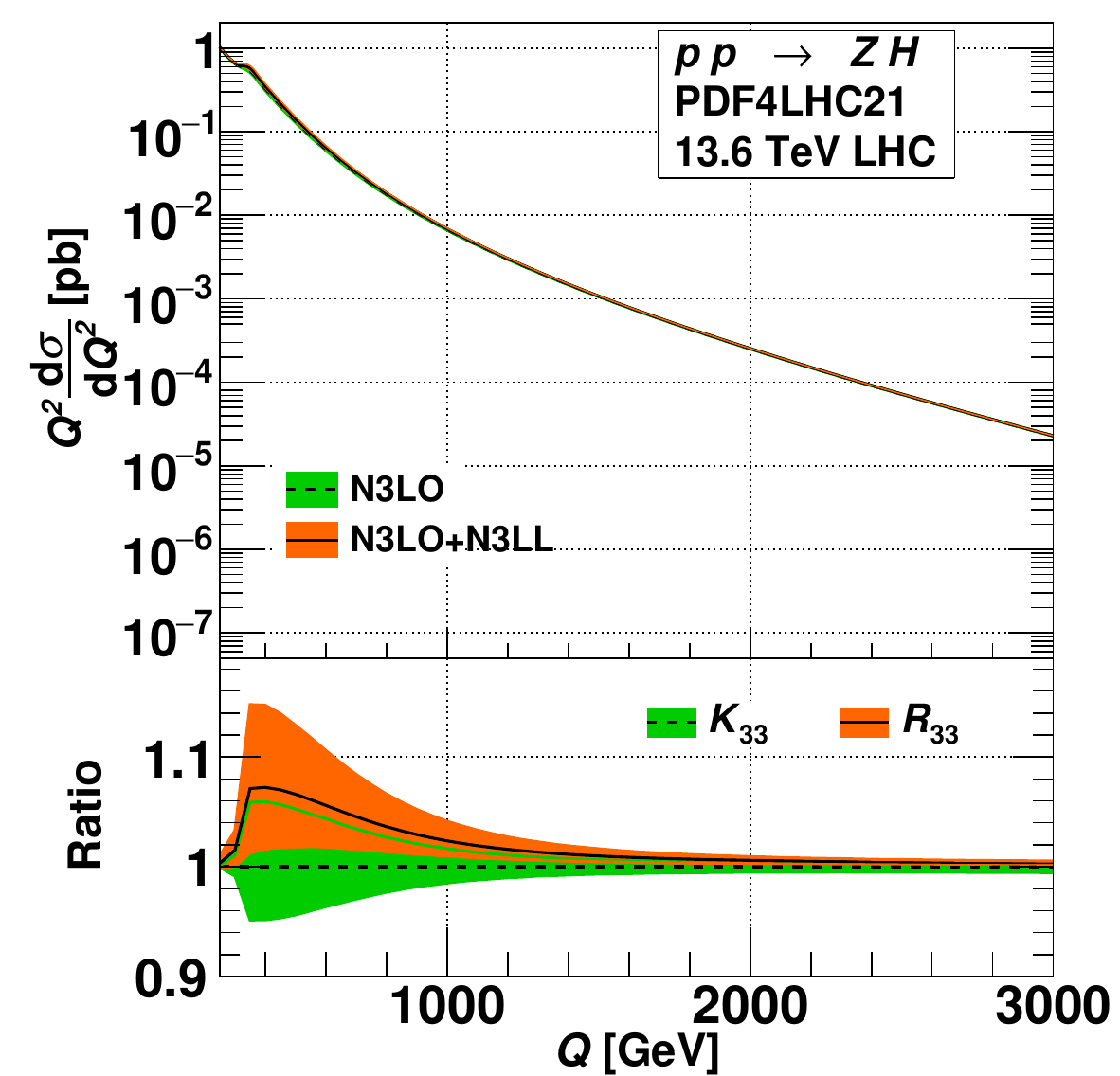}}
\vspace{-2mm}
\caption{\small{Different subprocess contributions with the unphysical scale uncertainties are presented here for both fixed-order and resum results.}}
\label{fig:dis_DY_gg}
\end{figure}

\begin{figure}[ht!]
\centerline{
\includegraphics[width=7.5cm, height=7.5cm]{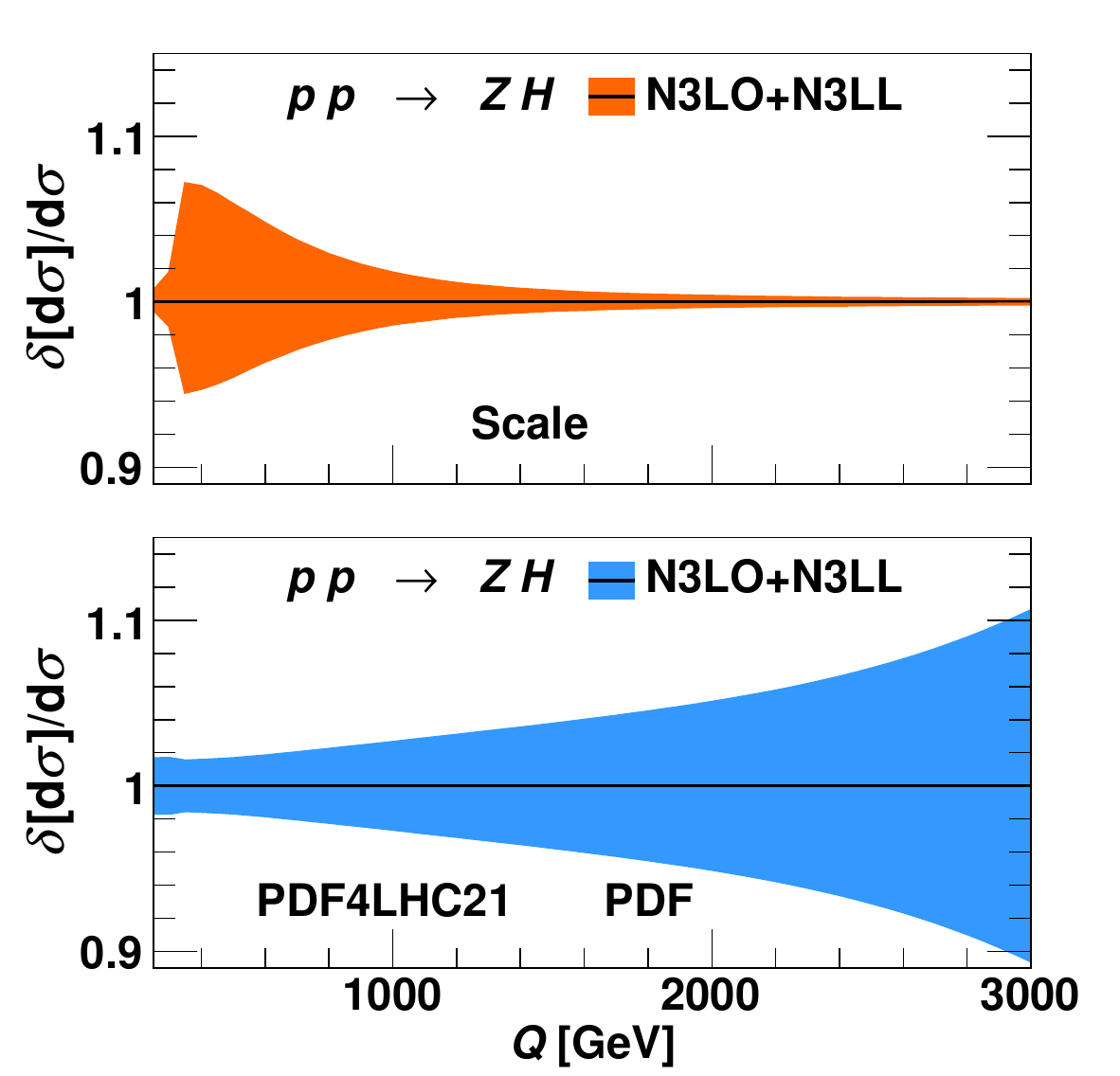}}
\vspace{-2mm}
\caption{\small{The 7-point scale uncertainty (upper panel) and the PDF uncertainty (lower panel) are shown for 
	the $pp \to ZH$ process at $13.6$ TeV LHC. }}
\label{fig:scale_ZH}
\end{figure}

Now, to obtain full $pp \to ZH$ results at the $\mathcal{O}(\alpha_s^3)$ level, we combine the gluon fusion results with other subprocesses for $ZH$ contribution. For the fixed-order case, we combine the contribution of the DY-type channel with the gluon fusion channel at $\mathcal{O}(\alpha_s^3)$. For the resummation case, we chose the standard SV resummed results. We combine the fixed-order results with SV resummed results for both DY-type and gluon fusion 
channels. The required N3LO+N3LL SV resummed results for the DY-type channel are obtained from \cite{Das:2022zie}. In the left panel of \fig{fig:dis_DY_gg}, we show these results separately for both DY-type and gluon 
fusion channels along with the scale uncertainties. The right panel of \fig{fig:dis_DY_gg} shows the combined results for invariant mass 
distribution at N3LO and N3LO+N3LL. The bottom panel highlights the 
enhancement from N3LO through the $K_{33}$ and $R_{33}$ factors. 
In \fig{fig:scale_ZH}, we show the scale and PDF uncertainties for the combined results at N3LO+N3LL.
We observe PDF uncertainty increases with invariant mass ($Q$) on the other hand theory uncertainty is the largest at the top-quark threshold
region and decreases with invariant mass.

%\subsection{Total cross section}
%We also studied the total cross section for the SV and NSV resummation and the results are presented in the \tab{tab:tableggZH}. We also studied theoretical uncertainties arising from the standard seven-point scale variation, the choice of nonperturbative PDF sets, and the variation of the strong coupling constant $\alpha_s(M_Z)$. 
%Additionally, we also present the combined uncertainties from PDF and $\alpha_s$ by adding the respective errors in quadrature.

\section{Summary}\label{sec:conclusion}
In summary, we have studied the impact of soft-gluon resummation on $ZH$ production in the gluon-fusion channel at the LHC up to next-to-leading logarithmic order for both SV and NSV levels. For the fixed order NLO computation, we use Born-improved NLO results. The NLO corrections contribute as 
large as $100\%$ of LO for the total $ZH$ production cross section in the gluon
fusion channel. We observed that the SV (NSV) resummation contributes an additional $19.4\%$ ($35.3\%$) at the NLL 
($\overline{\rm NLL}$) level over NLO in the low-$Q$ region. In the high invariant mass region (around $Q=3000$ GeV), the
SV(NSV) resummation reduces the seven-point scale uncertainties at the NLO level by a few per cent $5.0\% (1.4\%)$. We also quantified all sources of uncertainty in our computation. For experimental analysis, we combined the contributions from different subprocesses, including the soft gluon 
resummation effects, and presented comprehensive results for the invariant mass distribution at the LHC.

% \section*{Acknowledgements}
% The research of G. D.\ is supported by the Deutsche Forschungsgemeinschaft (DFG, German Research Foundation) under grant  
% 396021762 - TRR 257 (\textit{Particle Physics Phenomenology after Higgs discovery.}).
% The research work of M. C. K.\ is supported by the SERB Core Research Grant (CRG) under the project CRG/2021/005270. 
% The research work of K.S. is supported by the Royal Society (URF/R/231031) and the STFC (ST/X003167/1 and ST/X000745/1).
% \vspace{-10cm}

\end{document}